# Systems approaches and algorithms for discovery of combinatorial therapies

**Jacob D. Feala[1], Jorge Cortes[2], Phillip M. Duxbury[3], Carlo Piermarocchi[3], Andrew D. McCulloch[4], Giovanni Paternostro[1,4]**

[1]Burnham Institute for Medical Research, 10901 North Torrey Pines Road, La Jolla, CA 92037
[2]Department of Mechanical and Aerospace Engineering, University of California, San Diego, 500 Gilman Drive, 0411, La Jolla, CA 92093-0411
[3]Department of Physics and Astronomy, Michigan State University, East Lansing, MI 48824
[4]Department of Bioengineering, University of California, San Diego, 9500 Gilman Drive, 0412, La Jolla, CA 92093-0412

Author emails:

JDF: jfeala@burnham.org

JC: cortes@ucsd.edu

PMD: duxbury@pa.msu.edu

CP: carlo@pa.msu.edu

ADM: amcculloch@ucsd.edu

GP: giovanni@burnham.org

ABSTRACT

Effective therapy of complex diseases requires control of highly non-linear complex networks that remain incompletely characterized. In particular, drug intervention can be seen as control of signaling in cellular networks. Identification of control parameters presents an extreme challenge due to the combinatorial explosion of control possibilities in combination therapy and to the incomplete knowledge of the systems biology of cells.

In this review paper we describe the main current and proposed approaches to the design of combinatorial therapies, including the empirical methods used now by clinicians and alternative approaches suggested recently by several authors. New approaches for designing combinations arising from systems biology are described. We discuss in special detail the design of algorithms that identify optimal control parameters in cellular networks based on a quantitative characterization of control landscapes, maximizing utilization of incomplete knowledge of the state and structure of intracellular networks. The use of new technology for high-throughput measurements is key to these new approaches to combination therapy and essential for the characterization of control landscapes and implementation of the algorithms.

Combinatorial optimization in medical therapy is also compared with the combinatorial optimization of engineering and materials science and similarities and differences are delineated.

# 1. INTRODUCTION

Combinatorial Biological Control is a growing field that is benefiting from advances in systems biology, targeted therapeutics, and high-throughput biological measurement technologies as well as from new and established approaches from mathematics, physics and engineering.

The study of Combinatorial Therapies is the fastest expanding sub-discipline within this field, though therapeutic applications are not the only uses of the new principles and methods that are being discovered. Combinatorial approaches can also be used to optimize the survival and differentiation of cells *in vitro*, in synthetic genomics, to delay aging, and to improve physiological performance. Additionally, work in this field can help elucidate the strategies nature uses for combinatorial control and optimization at different scales, from evolution to organismal function. While until recently combinatorial therapies were based on largely empirical methods, new insights are arising from systems biology and from the integration of several biological and non-biological disciplines and are providing the prospect of more rational approaches to the therapy of complex diseases.

In this review, we describe the state of the art of combinatorial optimization of medical therapy, starting with relevant contributions from systems biology. We then compare it with more established techniques of systematic optimization in engineering and material sciences.

# 2. BIOLOGICAL ROBUSTNESS

The problem of controlling a biological system is complicated by the presence of active mechanisms to maintain function (whether homeostatic or dynamic, e.g., oscillations, differentiation) under the influence of environmental perturbations. This robustness is a desirable property of many complex systems that has been studied extensively for engineered systems and more recently applied to biological systems. Kitano [1] provided a mathematical framework to define and understand biological robustness that takes into account the maintenance of function for all possible perturbations, the probability of occurrence for each perturbation, and the limited resources available. Kitano also suggested that robustness is conserved for similar biological systems, meaning that under resource constraints there is some tradeoff in improving robustness to certain perturbations while becoming increasingly fragile to others [1]. Stelling *et. al.* [2] agreed with this conjecture, and a similar result was also a consequence of the theory of Highly Optimized Tolerance (HOT) for complex designed systems [3, 4]. The HOT theory assumes that the final products of biological evolution have similar robustness properties to designed systems in engineering.

It has recently been pointed out that biological networks, in order to optimize the crucial tradeoff between robustness and sensitivity [5] must have maximal mutual information exchange between nodes [6]. This su ggests that natural selection has acted in cells, neural systems, and other tissues so as to maximize the information transfer between signaling and receiving nodes across their networks.

Following this information theory approach, Kitano [1] also suggested the concept of controlling biological systems by applying the general principle of *spread spectrum communication* [7], in which selectivity and tolerance to noise is achieved by spreading the signal through many channels or nodes in a network. In the case of drug design, this approach suggests that efficient and highly-selective therapies could be obtained combining a large number of low-dose compounds. This is supported by numerical studies showing that multiple weak hits can affect a regulatory network more efficiently than a complete inactivation of selected nodes [8].

The consequence of biological robustness in terms of our attempt to control these systems is that specific targets for robustness and fragility must be understood in order to circumvent the intrinsic tendency of biological systems to maintain function (whether in the normal or diseased state) under external influences [1]. Theoretical and applied examples of distributed robustness and redundancy in the face of cascading failure support the idea that biological networks are robust to single interventions [9-12]. Therefore, multiple drugs will be needed to address robust disease states.

3. THE PROBLEM OF COMBINATION THERAPY

It is indeed becoming increasingly evident to the clinician treating a complex disease or to the scientist studying a complex biological network that accurate control is more likely to be achieved by using multiple interventions. Since therapeutic molecules are increasing in specificity (as in the case of targeted drugs), and since our knowledge of the complexity of biological networks is advancing, it is becoming more feasible to consider drugs not as remedies for specific disorders but rather as a kit of molecular tools that can be combined for specific therapeutic purposes. Additionally, as we do when evaluating the health effects of natural products (for example, red wine or green tea), it may be acceptable to consider a novel large combination of compounds as a whole rather than studying all its components separately.

Because drug effects are dose-dependent, several doses need to be studied and, when therapeutic interventions on multiple targets are necessary, the number of possible combinations rises very quickly (this problem is often referred to as combinatorial explosion). For example, if we were to study all combinations of 6 out of 100 compounds (including partial combinations containing only some of these compounds) at 3 different doses we would have

$$\Sigma^{6}_{j=1} \, Binomial(100,j)*3^{j} = 8.9*10^{11}$$

possibilities. This example suggests that the problem will require a qualitatively new approach rather than more efficient screening technology alone. Many cancer chemotherapy regimens are composed of 6 or more drugs from a pool of more than 100 clinically used anticancer compounds.

The following quotation from the final chapter of a recent textbook on the Biology of Cancer [13] by Robert Weinberg illustrates the growing awareness of this problem and of new opportunities for its solution:

*"Traditionally, new drugs have been evaluated as single agents during pre-clinical development and Phase I clinical trials. This practice contrasts with the growing belief of cancer researchers that most monotherapies are unlikely to yield curative treatments and that, with rare exceptions, truly successful clinical outcomes will depend on the use of combinations of anti-cancer drugs. … At present, the choice of drugs to be used singly or in combinations is inspired by biological intuition or poorly performed guesses. Increasingly over the coming decade, strategies for organizing multi-drug treatment protocols will be influenced by our rapidly evolving understanding of the design of the signaling circuitry within human cells and by molecular diagnostics that tell us how certain signaling pathways have been perturbed in some tumors and not in others."*

A similar view on the limitations of the traditional approach to drug combination has been offered by Sridar Ramaswamy in a recent Commentary in The New England Journal of Medicine [14]. Other reviews stressing the need for a more systematic approach to combination therapy have been published by Dancey and Chen [15], Hopkins [16] and Zimmerman et. al. [17]. A recent editorial [18], commenting on the disappointing results of a clinical trial of combination therapy for colorectal cancer [19], suggested that combining the new targeted therapies might be even more challenging than combining cytotoxic chemotherapies, because of subtle interactions in intracellular signaling.

4. COMBINATORIAL OPTIMIZATION IN MEDICAL THERAPY

In the following sections we describe the main classes of strategies for designing drug combinations (or closely related approaches) presented in the literature.

**Class A - Empirical method (not systematic)**

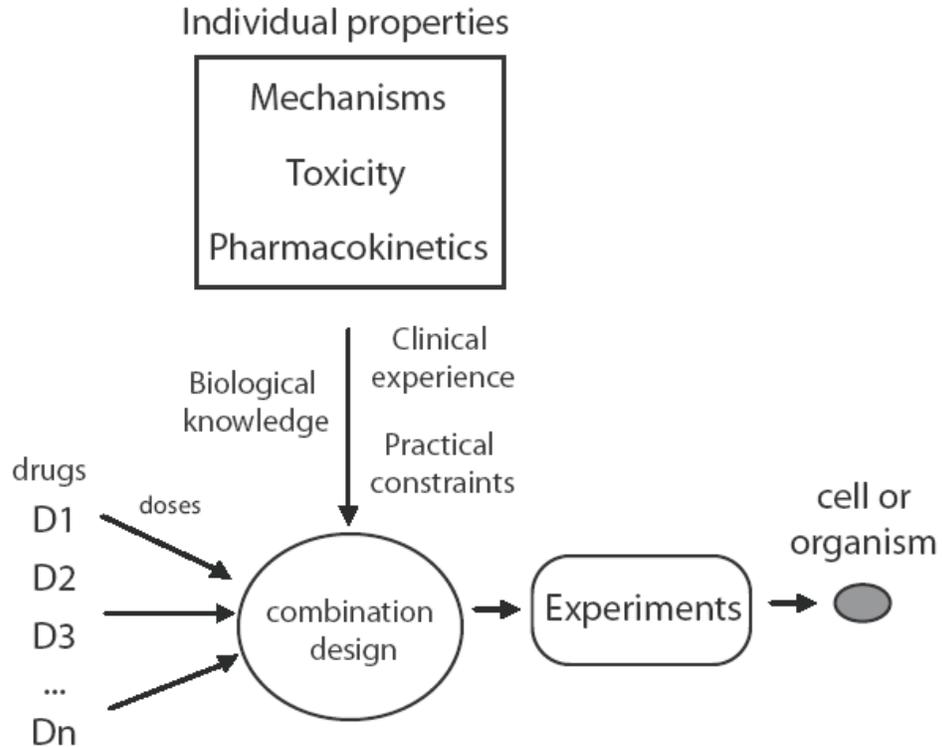

**Figure 1: The empirical method.** This approach is non-systematic, instead designing combinations based on the clinical experience of doctors, knowledge of biological mechanisms, and practical constraints in the design of clinical trials.

In this class we include the traditional approaches. It is important to stress that by empirical [15, 20] we mean that the approach is not systematic. We do not mean that empirical approaches are non-scientific: on the contrary, many clinical trials of drug combinations are very rigorous and clearly useful [18, 19]. A common assumption in this class is that only drugs that are effective individually should be used as part of a drug combination [21]. This assumption might be related to FDA requirements, which however now seem to be evolving [13]. Because clinical trials are very expensive and many are sponsored by drug companies, it has been pointed out that commercial factors play a role in the choice of which combinations are tested [22]. The FDA regulates which individual drugs can be prescribed but individual physicians are free to use their clinical judgment when prescribing them in combination. We do not assume that a systematic approach would be necessarily be superior, but randomized, double blind, prospective trials could compare the efficacy of specific systematic approaches with physician choices. Improving the evaluation of the efficacy of medical therapy is one of the measures suggested for reducing inefficiencies in health care expenditures [23].

## Class B – Brute force (exhaustive enumeration)

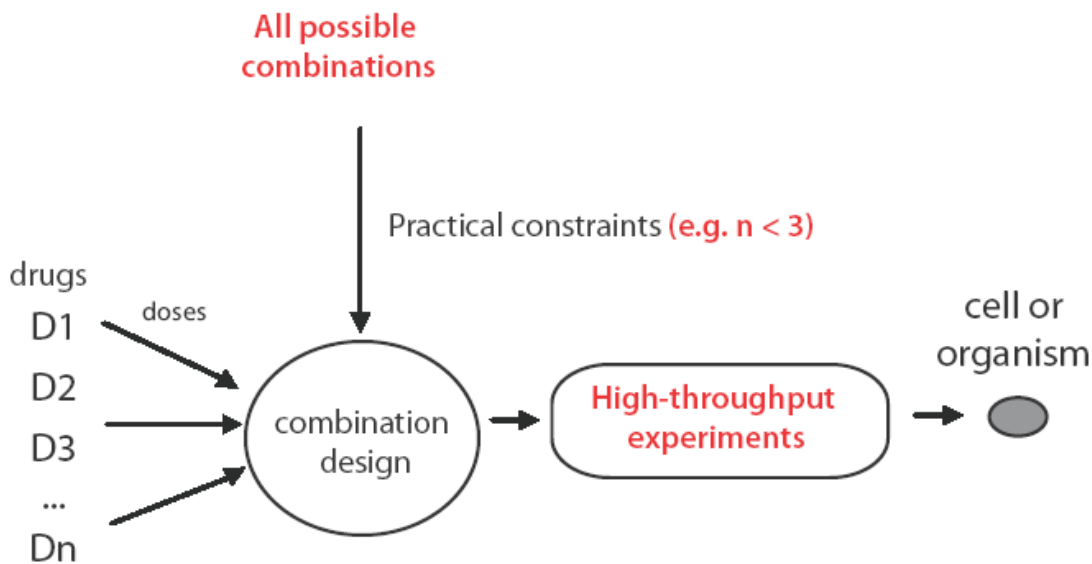

**Figure 2: The brute force approach.** "Brute force" is a term meaning exhaustive testing of all possible combinations. High-throughput screening technology allows the testing of pairs of drugs over a range of doses, but combinatorial explosion usually prevents exhaustive measurement of combinations of more than 2 drugs except when exploring combinations from very limited sets of compounds.

This approach has been called "brute force" to indicate that all combinations up to a certain size are tested [17, 24]. Because of the mentioned rapid combinatorial explosion this means that only combinations of a small size can be studied. A "brute force algorithm" is a technical definition for an algorithm that performs an exhaustive search. It does however require a considerable amount of judgment to decide which set of drugs to test and which biological assay to use.

Most studies in the literature only study pairs of drugs, and it would be very difficult to extend this approach to larger combinations unless the set of candidate compounds to be used is small. Greco [25] reviews the traditional literature on drug combinations and drug synergy, which was mainly developed to study interactions of two drugs. Te Dorsthorst [26] presents a recent application of this approach applied to the study of antifungal agents. A more systematic approach is suggested by Borisy *et. al.* [27] using high-throughput screening in several different biological systems and measuring hundreds of thousands of pair-wise drug interactions.

The main idea underlying this approach is that unexpected therapeutic activities, not predictable from single drug studies, might manifest themselves when multiple drugs are used, and this seems to be the case, see e.g., [18, 19, 27].

**Class C – Statistical association**

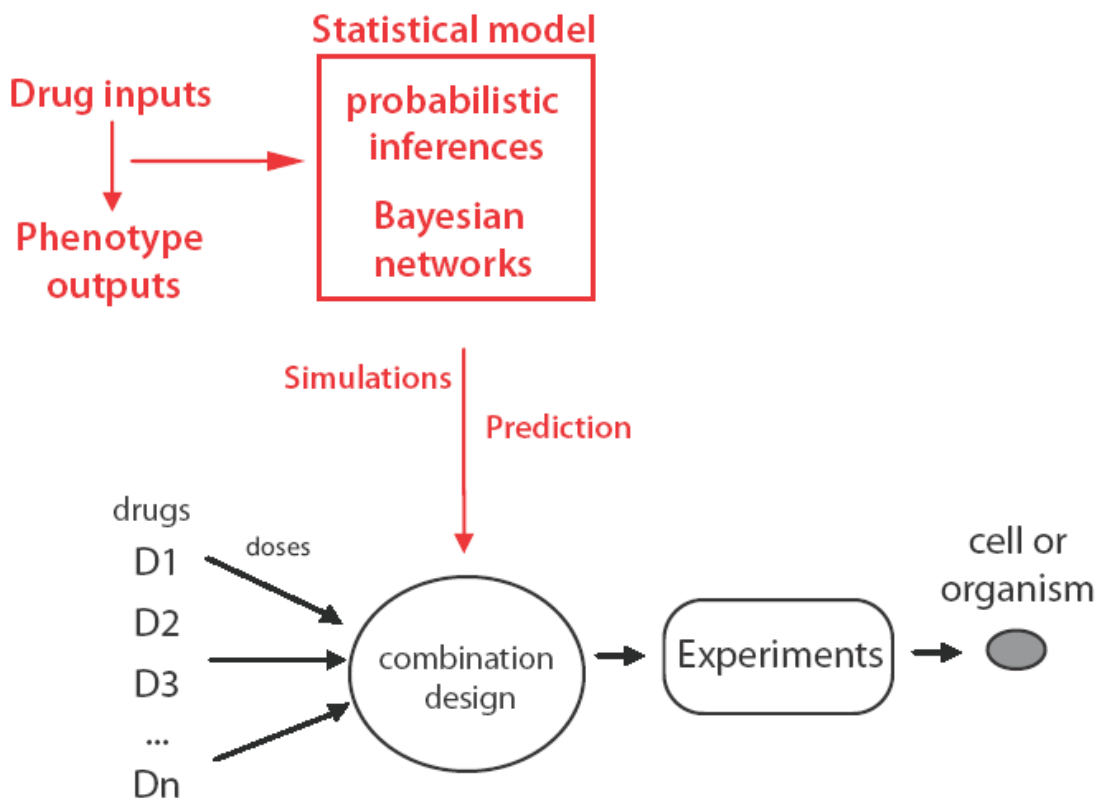

**Figure 3: Statistical association models.** This approach attempts to design combinations based on the desired phenotype and linear combinations of input/output relationships of single drugs.

Statistical models can be built based on correlations between drugs and phenotype outputs. These models do not rely on an explicit characterization of the biological networks but treat the system as a black box. By predicting phenotype outputs in response to new sets of drug inputs and assuming simple, usually linear, relationships, these statistical models can aid in the design of drug combinations.

Lamb *et. al.* [28] compiled a database of gene expression profiles in response to single agents. The "Connectivity Map" allows new samples to be matched to gene expression patterns so that the best set of single drugs can be found. Although the database can be used to connect related small molecules, using data from a limited number of cell lines and compounds, no attempt has been made to use the resource to design new combinations. Janes *et. al.* [29] used partial least-squares regression to develop a statistical model linking a signature of apoptosis markers to an array of activity metrics in 19 signaling proteins, when activated by different combinations of cytokines. Their method demonstrated the ability of the statistical model to determine a canonical basis set for combinatorial control of the apoptosis response, using cytokines as inputs.

Taking advantage of the large-scale screen of drug responsiveness in the NCI-60 panel of cancer cell lines, as well as microarray data for each cell line, Potti *et. al.* [30] derived gene expression signatures to classify cell lines on the basis of sensitivity to various chemotherapeutics. These expression signatures could predict drug sensitivity in cell lines as well as in patients in clinical trials. When individual expression signatures were combined to produce a chemosensitivity signature for a specific drug combination (TFAC – paclitaxel, 5-FU, adriamycin, and cyclophosphamide), these genes produced a statistically significant distinction between responders and non-responders in a retrospective analysis of clinical trials for combination efficacy [30].

It is very hard for pure statistical correlations to capture the non-linearity, complexity and variability of biological networks, but mechanistic network models are still incomplete and the parameters are often not well known, making the statistical approach a useful tool especially when dealing with large omic datasets.

**Class D- Explicit model-based methods**

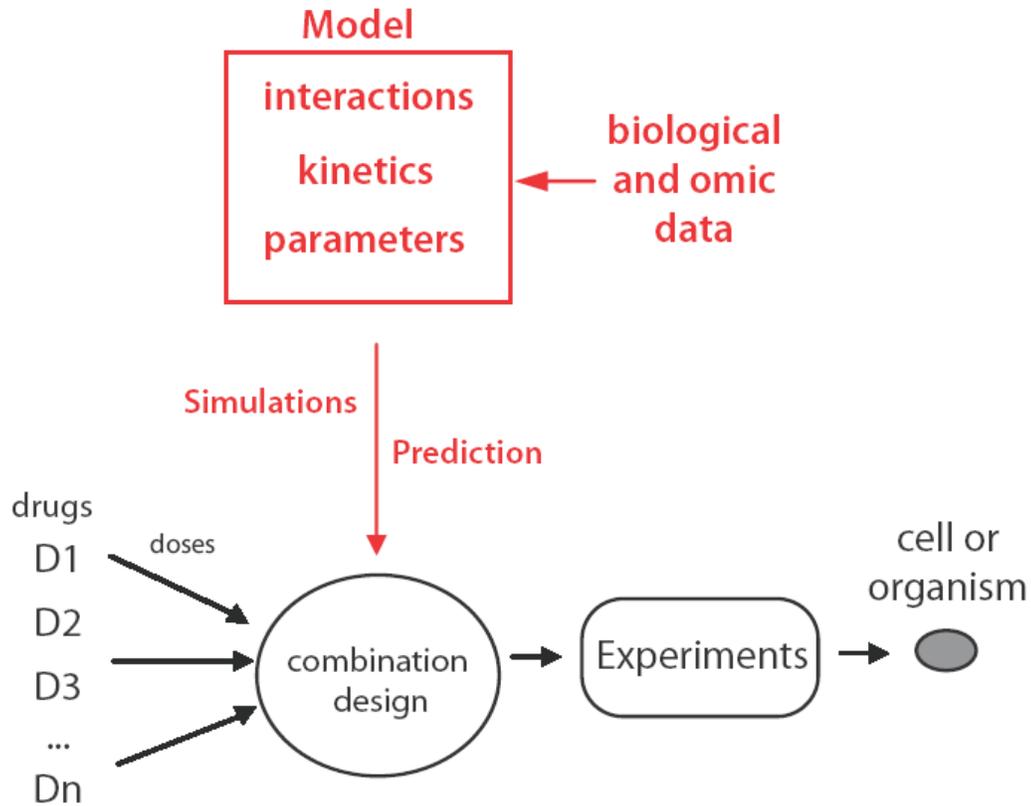

**Figure 4: Model-based approach to designing combinations.** Explicit models of biochemical interactions, usually fitted to measured biological data (including genome-wide or "omic" data such as microarrays), are used to predict optimal combinations via simulation.

In this class, biological measurements (either obtained by the authors or derived from the literature) are used to build explicit models of the target biological network for the optimization of drug combinations using simulations. Any control system implicitly assumes a model of the system being controlled [31], but in an explicit model there is a direct correspondence between the elements of the model and reality.

Araujo *et. al.* [32] used a medium-size system of ordinary differential equations to model a handful of combinations on the EGFR network and found synergistic interventions. Calzolari *et. al.* [33] developed a model of the apoptosis network and identified general strategies and potential targets for combined drug interventions to obtain selective cell death. Yang et. al. [34] modeled the arachidonic acid metabolic network, related to inflammation, with a 2-step simulated annealing approach to optimize (1) drug candidates and (2) levels of intervention for multiple targets. Hua and Palsson [35] is one example of a large body of work on control and analysis of metabolic networks using genome-scale constraint based approaches. They predict and design gene knockouts to optimize metabolite production.

Tvieto and Lines [36] took advantage of well established mechanistic models of cardiac myocyte electrophysiology [37] to perform an *in silico* search for combinations of ion-channel targets that minimized the effects of ischemia on action potential morphology [36]. They confirmed that the optimal combination therapy reduced the pro-arrhythmic consequences of regional ischemia in a spatially coupled model of myocardial tissue. The method made use of the well known Nelder-Mead optimization method [38] to find an optimal "drug vector" describing the fractional inhibition or agonism of six channel conductances. The optimization minimized a distance function that measured time integrals of squared differences in membrane potentials and current densities between treated and untreated normal cells, and between treated ischemic cells and normal cells.

Some authors have the opposite but related aim of using drug combinations to build network models. The focus is on system analysis rather than on system control. Lehar *et. al.* [39] used response surfaces of pair-wise combinations to predict network motifs surrounding the drug targets. The shape of a simulated dose response surface can distinguish parallel and serial conformations, branching, feedback, etc. in a generic computational abstraction of a metabolic network. They then tested the method on the well-known sterol metabolism pathway in yeast, noting that the response surfaces of drug pairs correctly predicted the topology of the pathway. The method can be generalized to higher-order combinations as well.

Nelander *et. al.* [40] built a nonlinear multiple-input, multiple-output network model using experimentally measured drug pair inputs and phospho-protein outputs in MCF7 breast cancer cells. They applied Monte Carlo methods to optimize the choice of network connectivity, and a gradient descent technique to optimize parameters and interaction weights. As opposed to the response surface method of Lehar et. al. [39], this method produces globally parameterized computational models ready for simulations.

This class is likely to be an essential component of a successful approach to the optimization of drug combinations but we should not forget that we do not have the luxury of having a specific drug to modulate each node of a network. Targeted therapies are only available for a minority of proteins, and they normally have off-target effects, some of which are not known.

It is also possible to envisage that explicit network models and statistical associations could be combined, as it has been done in other fields of systems biology, for example in the use of microarrays and protein interaction networks for tumor classification [41]. The statistical approach could either be used to define the parameters of the model or directly to suggest combinations. We are not dedicating a separate section or figure to this class (which can be called class E) because we have not found direct examples in the field of combinatorial therapy. It is too early to exclude that these types of approaches might emerge, but if class E will remain empty or scarcely populated it might be useful to consider the reasons.

**Class F- Model free biological search algorithm**

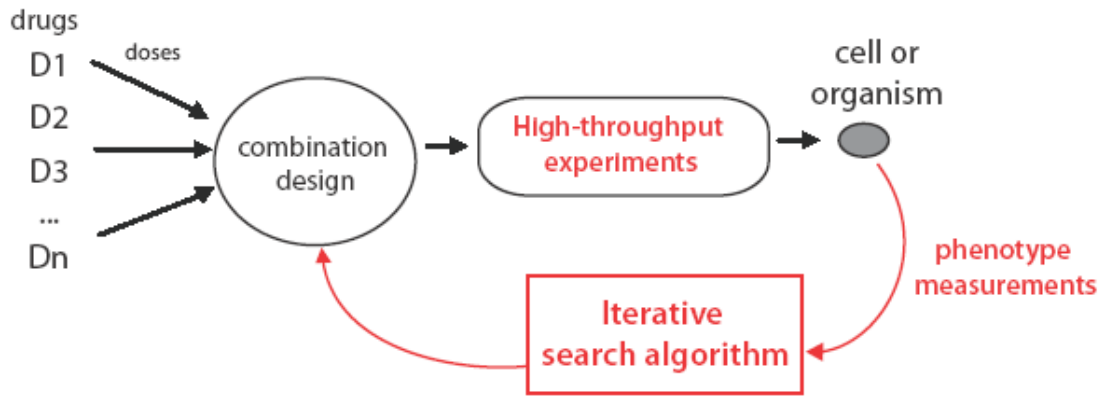

**Figure 5: Biological search algorithm.** In this model-free approach to combination design, drugs are iteratively combined and measured, in effect searching through the vast space of possibilities.

This class describes biological search algorithms, where the search is not conducted *in silico* but directly using biological measurements from *in vitro* or *in vivo* systems. These methods can be stochastic, for example using Monte Carlo or evolutionary algorithms, or non-stochastic.

Wong *et. al.* [42] developed a closed-loop control algorithm, using a microfluidic platform to implement an iterative stochastic search (Gur Game algorithm) for optimal drug combinations. The algorithm efficiently discovered potent combinations for inhibiting virus infection of fibroblasts, and was able to regulate NFκB levels by discovering optimal cytokine levels. Sun *et. al.* [43] extend this work by using the Gur Game algorithm followed by the Differential Evolution algorithm, a type of genetic algorithm that can be used for parallel search.

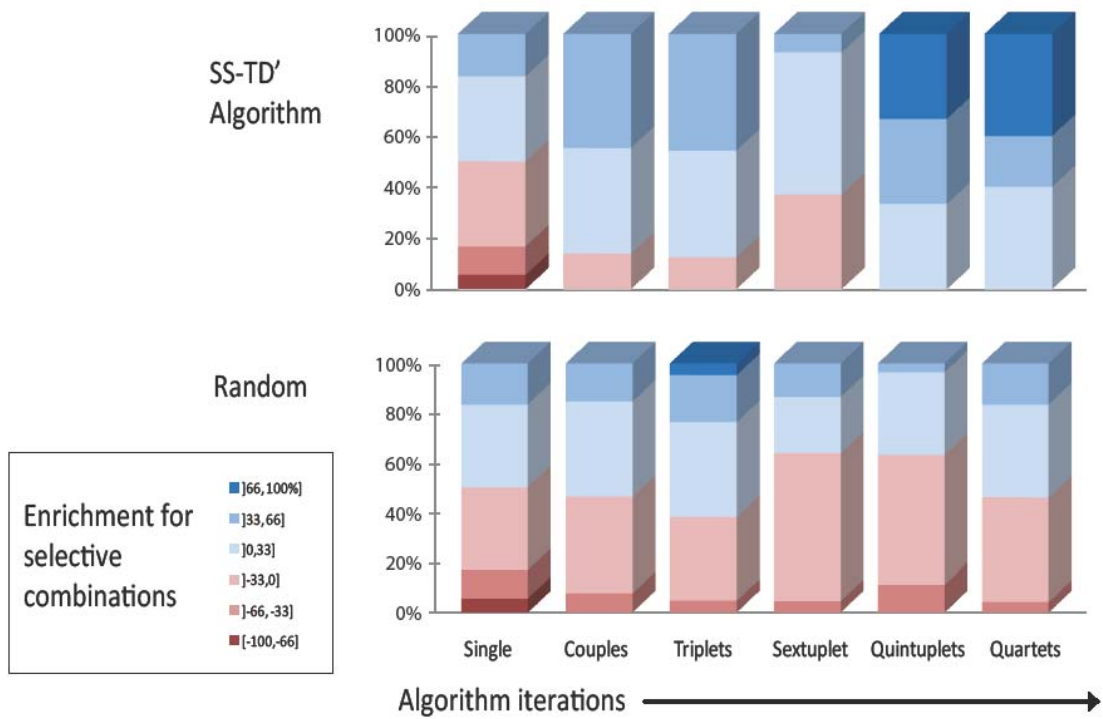

**Figure 6**: **Using the biological search algorithm ("Stack Sequential – Top Down" or SS-TD) to optimize combinations for cancer cell selectivity.** The colors indicate the selectivity of the drug interventions and the aim is to find treatments with high selectivity for one of the cell lines. This desired selectivity is shown as dark blue. The red shades are partially selective for the other cell line. Iterations of the algorithm apply different sizes of combinations, starting with individual drugs. A statistically significant enrichment of the desired selective combinations (dark blue) is shown. Adapted from [44].

Calzolari *et. al.* [44] used non-stochastic algorithms derived from one used in digital decoding, the stack sequential algorithm, to perform biological searches for drug combinations, both *in vivo* (using fruit flies) and *in vitro* (using cell lines – see Figure 6). This approach is a first example of application of information theory concepts and techniques to the control of biological systems.

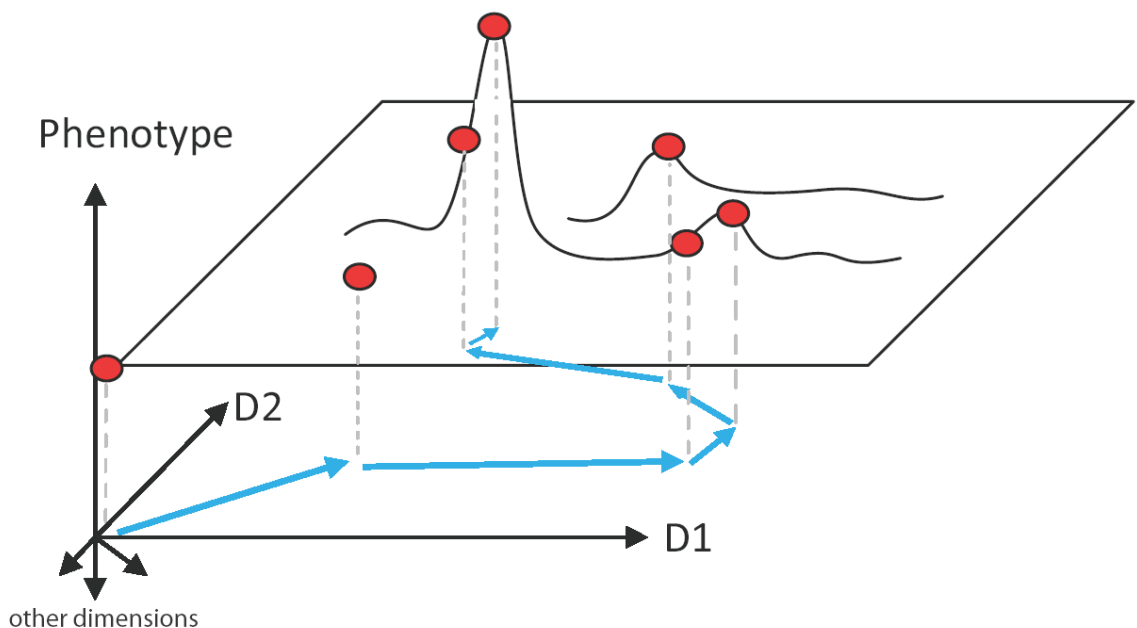

**Figure 7: Diagram of the iterative search framework.** At each iteration, a high-throughput assay measures cellular (or organism) phenotype in response to a drug combination, which is fed into the algorithm. The algorithm then generates new candidate combinations to test based on the previous results. Conceptually, this is equivalent to a search through the phenotype landscape in the space of possible drugs and doses.

Further progress within this class would benefit from the availability of a corpus of fully factorial (exhaustive) drug combination datasets that could be used to compare algorithms and to develop new ones. A detailed statistical analysis of these fully factorial datasets could help in testing and evaluating the efficiency of different optimal drug combination search strategies. This type of analysis can be carried out by characterizing the general properties of the *control landscape* of a given biological system (Figure 7).

The notion of *landscape* represents a very useful concept, commonly used in the analysis of many complex systems encountered in physics, biology, computer science, and engineering [45]. The analysis of the general properties of a landscape is of interest in the context of optimization problems since it provides hints on the type of optimization algorithms that are more likely to succeed (Figure 8). For instance, in the control of molecular systems by laser pulses, different choices of parameters lead often to equivalent outcomes. This is reflected in the fact that in these systems optimal solutions lie in high-dimensional continuous surfaces of the control landscape [46]. In such types of landscapes, which can be visualized as having plateau regions in the control space, an efficient optimization algorithm needs to "jump" between the different plateaux. This can be better achieved using stochastic algorithms, for example genetic algorithms (see Section 5: Combinatorial optimization in engineering).

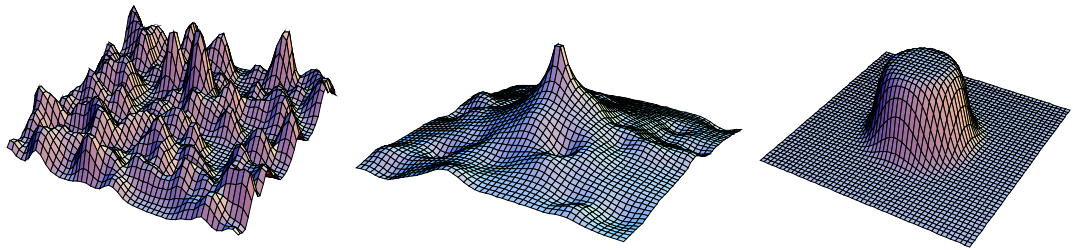

**Figure 8: Scheme of different types of landscapes.** The structure of the control landscape can inform the choice of the search strategy. Left: Rugged, found e.g. in spin glasses as a result of competition between ferromagnetic and anti-ferromagnetic interactions. Center: Funnel, found e.g. in protein folding fitness landscapes. Right: Robust, found e.g. in quantum control problems in which often control parameters give perfect control or no control at all.

## Class G- Model plus biological search algorithm (integration of C, D and F)

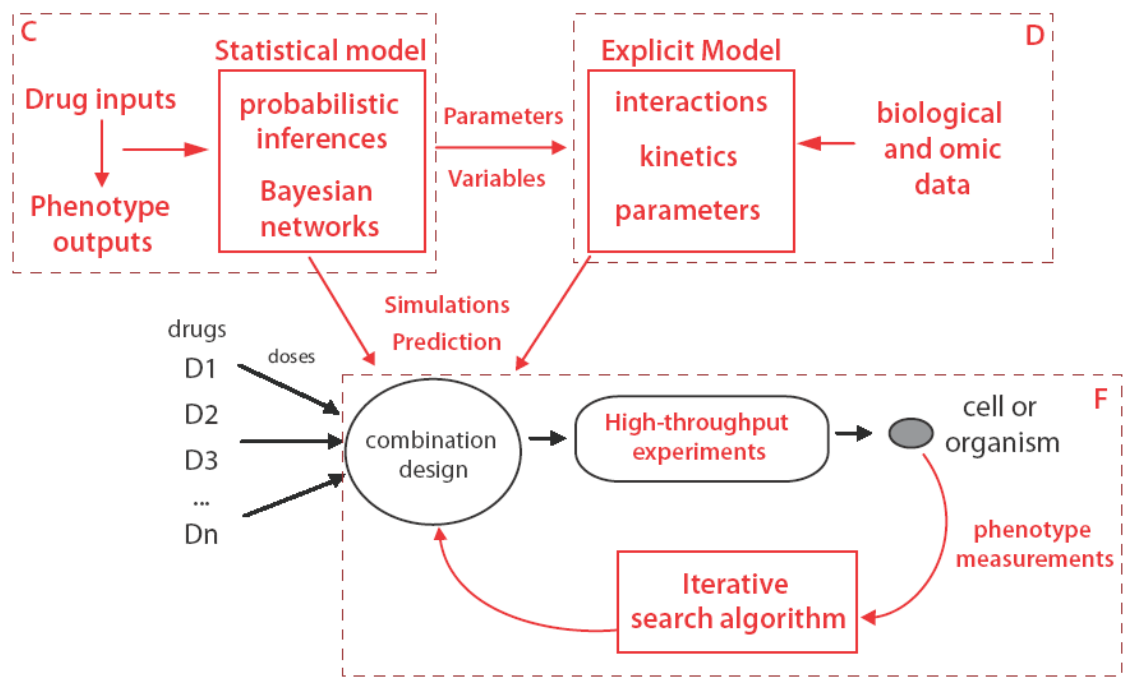

**Figure 9: Modeling approaches integrated with experimental search.** A more systematic approach could integrate information from statistical and explicit models. Model-based predictions of effective perturbations can be combined with closed-loop iterative experimental search to refine the drug combination.

Calzolari *et. al.* [44] discussed the motivation for this class and describe in detail algorithms that can be used to integrate biological search with statistical association and explicit model approaches. What characterizes this class is the combination of the experimental search of class F with in silico models (class C, class D or both).

The algorithms described by Calzolari *et. al.* [44] (see Figure 10 of that paper) can include the results of *in silico* models within the ranking used to guide the steps of the experimental search. They therefore do not simply combine the outputs of separate approaches but integrate them fully.

This class of strategies for combinatorial therapy could also utilize the vast amounts of genomic data originating from personalized medicine efforts. Sequencing individual genomes at an affordable cost might soon be possible. Personalized combinatorial therapy might therefore be based both on the individual genome (and other omic data) and on an experimental search (as in class F) using the individual's own cells.

As for class F, it would be desirable to obtain a corpus of fully factorial (exhaustive) drug combination datasets and corresponding omic data that could be used to develop strategies of this type.

5. COMBINATORIAL OPTIMIZATION IN ENGINEERING

The solution to many engineering problems incorporates ideas that can be traced back to combinatorial research, where the individual elements to be combined change greatly among different applications. Examples include the selection of waiting time combinations at airport gates in order to minimize air traffic congestion [47], the combined tuning of the parameters of several controllers (each one contributing to a desired objective) in supply chain management [48], the optimization of engine parameters such as intake, exhaust, and spark timings according to operating conditions to improve fuel consumption [49], or the adjustment of ion-intensity model parameters in magnetospheric image processing [50].

Exhaustive enumeration (as in class B) is a trivial but very general problem-solving technique in combinatorial optimization that consists of systematically enumerating all possible configurations in searching for the optimal solution. Because of combinatorial explosion the method is only suitable for low-dimensional problems.

In computational science and engineering, when a reliable model of the problem is available (as in class D), the search for optimal combinations of the individual components can be carried out with a variety of optimization methods. For example, in convex optimization, a known and convex objective function is to be optimized over a convex domain. For convex functions, any local minimum is also global. Also, the set of all minima is convex. Finally, if the function is strictly convex, then there exists at most one minimum. Powerful methods exist to obtain global optimizers in this case. These include interior-point methods, subgradient methods, elliposoid methods, and duality. Numerous problems that are not convex in their original variables can be transformed into convex form after a suitable change of variables, and are therefore amenable to convex optimization techniques [51].

In stochastic optimization, good models are often not available, or we do not have an analytical expression for the objective function that we wish to optimize, or the evaluation of this function is costly or only possible through experimentation. In such cases, methods that do not depend on direct gradient information or exact measurements become very useful. Recently, there has been an increasing interest in recursive stochastic optimization methods that are based on noisy measurements of the objective function to be optimized, rather than on direct measurements of its gradient. These algorithms have the advantage of not requiring detailed modeling information describing the relationship between the parameters to be optimized and the objective function. The establishment of rigorous theoretical guarantees on the convergence properties of these algorithms is a subject of current research. Overall, gradient-free stochastic optimization methods have similar convergence properties as gradient-based stochastic algorithms. Although the theoretical asymptotic characterization states that gradient-free methods typically require a larger number of iterations until convergence, they offer great computational savings in the number of function evaluations required step and are more robust to noise and model uncertainty. A method that seems particularly well suited for combination therapy (in class F) is the "simultaneous perturbation stochastic approximation (SPSA)" [52]. The SPSA is based on approximating the gradient of the objective function using only two (possibly noisy) measurements, regardless of the dimension of the optimization problem. This feature allows for a significant decrease in the implementation cost in high-dimensional optimization problems. The applications of SPSA include model-free control [53], simulation-based optimization, and image processing. The global convergence properties of SPSA have been recently studied in [54].

Strategies inspired by evolution are potentially useful in class F. Evolution strategies use mechanisms inspired by biological evolution such as mutations, crossovers, and selection. Examples include genetic algorithms, particle swarm optimization, and ant colony optimization. In an evolutionary algorithm, candidate solutions play the role of individuals in a population. At each iteration, the fitness of every individual is evaluated. Multiple individuals are stochastically selected from the current population based on their fitness and modified according to the biological mechanisms mentioned above to form a new population. The new population is then used in the next iteration of the algorithm. The implementation of these algorithms therefore requires two basic elements: a representation of the solution domain and a fitness function to evaluate potential solutions. The fitness function is determined by the objective function of the optimization problem. In general, no convergence guarantees are available for these methods.

Simulated annealing simulates the cooling process of a physical system. Starting at a very high temperature, a series of temperature decreases is performed. At each temperature step, the state of the system evolves according to some transition probabilities chosen according to the physical problem. The idea is that, at high temperatures, movements that result in temporary increases of the objective function are more likely to happen, while as the temperature approaches zero, disappear in favor of a steepest descent motion. If the cooling of the procedure is performed slowly enough, then, with high probability, the global optimum of the optimization problem is achieved [55].

Search algorithms potentially belonging to class F are typically used in engineering within a model of the system, rather than experimentally in the real system. Research in materials is a distinct exception to this observation, as we discuss below.

An analogy with class G can be found in robotics and artificial intelligence. Explicit models of reality are not sufficient for robots to manage a complex environment [56]. An alternative approach that started from simpler stimulus-response algorithms was more successful and was later integrated with the older models in hybrid architectures [56]. This process is similar to the evolution of the nervous system which is based on stimulus-response mechanisms of increasing complexity in lower invertebrates, integrated (but not replaced) by representations of reality within the brain of higher organisms. Compare Figure 9 for class G with Figure 1 of [57].

6. COMBINATORIAL OPTIMIZATION IN MATERIALS RESEARCH

Modern combinatorial materials research is considered to have commenced in 1995 [58] leading to a rapidly expanding literature that according to a recent estimate is over 10,000 publications [59]. Experimental methods have been developed for combinatorial screening of a wide range materials systems [58-60], including: catalysts [59]; thin polymer films [59, 61]; ternary inorganic compounds [58]; and polymers for biomedical applications [62]. High throughput cell based assays with over 1000 units [63] have been used to assess the effect of polymers on cellular response [62-64]. There is a strong industrial impetus for these developments and national centers of excellence in combinatorial materials research, including the NIST center for combinatorial methods (NCMC), and the Combinatorial Sciences and Materials Informatics Collaboratory (CoSMIC) at Iowa State, have been established. Very large combinatorial arrays have been fabricated, particularly in the area of inorganic films where screens containing thousands of different stoichiometries have been reported [58].

Most combinatorial materials experiments operate in a cycle where the early experiments generate interesting regions of parameter space, while later experiments focus more closely on leads generated in previous cycles [59, 60, 65, 66]. Lead identification has also been enhanced by using statistical indicators based on prior knowledge and by incorporating theoretical insights. Use of prior knowledge through quantitative structure-property (QSPR), structure-activity (QSAR) or composition-activity (QCAR) relations profits from machine learning procedures and through use of artificial neural nets [59]. Optimization methods, such as Monte Carlo and genetic algorithms, have been used to assist in the choice of promising leads and hybrid methods incorporating prior knowledge through machine learning and optimization methods are emerging (see Figure 8 of ref. [59]). The methodologies used in combinatorial materials research have a great deal in common with those used in searches for drug combinations, and the cross-fertilization of these areas is promising.

In some cases theoretical models are sufficiently accurate that a purely computational approach to combinatorial materials design is feasible, as illustrated by use of density functional or quantum chemistry methods with optimization methods to predict new compounds or chemicals with desired properties; for example semiconductors with desired band gaps [67], new stable alloys consisting of four elements [68], and molecules with optimized properties [69]. *De novo* computational methods for designing individual drugs are also developing rapidly [70].

Approaches to combinatorial materials research can be classified according to the categories used in this review, ranging from more empirical and brute force experimental approaches [58, 62] (as in classes A and B), to the use of statistical association methods [66] (as in class C), methods based on explicit models [67-69] (as in class D), the use of search algorithms based on the experimental data [59] and often using data mining methods [59, 60] (as in class F). Recent approaches strive toward combining all these methods in meaningful ways [59, 71] (as in classes E,G).

7. FUTURE DEVELOPMENTS

As described in Steil *et. al.* [72] and in Weinzimer et. al. [73], it is possible to dynamically control in vivo, with real time feedback, the effect in humans of a single drug (in this case insulin). In the engineering literature the word control is usually used in this dynamical sense. The previously cited biological references have the more limited aim of optimizing drug combinations to be applied at a single time point. This class of feedback controlled *in vivo* algorithms, which would most likely include all previously described approaches and probably others, seems to us one of the possible future developments of combinatorial biological control using drugs. The success with a single drug shows that this expectation is not unrealistic.

The choice of a clinically relevant biological output, to be optimized in the cells or in the organisms, is also a difficult problem that is shared by scientists working on individual drugs. Not only can we expect increased knowledge from omic data to help translate from experimental systems to patients, but the emergence of adaptive designs in clinical trials might also allow these to be structured as search algorithms [74].

A more systematic approach is also needed for choosing the subset of drugs within which to search for optimal combinations. Systems biology can probably assist this choice, especially for the increasing number of targeted therapies available [75, 76], where a more precise correspondence between drug and biological target can be established.

We have described many possible algorithms for efficiently searching the space of possible drug combinations, and this supports the need for a corpus of fully factorial drug combination datasets (landscapes) that could be used by interested scientists to test and compare different algorithmic alternatives. Beside the drug combinations and the biological outputs, the datasets should also contain omic profiles (as genomics, microarray, proteomics and metabolomics) to specify the state of the model variables, at least for the individual drugs and for the main maxima in the landscape.

Fundamental experimental and theoretical understanding of the morphology of the control landscape will assist in developing improved algorithms to direct combinatorial therapeutic search in favorable directions. In hard computational problems it is well known that as the degree of constraint increases, there is a phase transition from problems where it is relatively easy to find satisfying solutions to a phase where it is difficult to find good solutions. This phase transition has been studied intensively in model complex systems such as random K-SAT [77], however the implications for combinatorial drug discovery remain to be explored.

Such new datasets may spur cross-fertilization from other quantitative sciences and new developments in computational biology, in the same way that large-scale genomic and proteomic datasets have already done [78].

8. CONCLUSIONS

A systematic approach to the field of combinatorial drug interventions has only recently started and it is too early to have a complete picture of all differences versus combinatorial optimization in material sciences or engineering. There are, however, a few aspects we would like to point out.

When explicit models are used in non-biological optimization efforts, there is usually just one type of model used in a particular application. Calzolari *et. al.*, however, showed (Figure 10 of [44]) separate models of metabolism and of signal transduction as possible sources of inputs for the search. It would not be easy to unify them, even if the outputs can be combined. Other relevant classes of biological models with distinctive features do exist, for example electrophysiology models. Including different heterogeneous models might be especially necessary in the optimization of drug combinations because of the partial coverage of these models and because of the complexity of biological systems.

Another feature to be noted is that the relation between drug and target is not simple; even targeted therapies have some off-target effects and we do not have a drug for every protein. It is therefore not possible to tune selectively all the parameters of a biological model as it might be done in many engineering applications.

Finally we must consider that variation is an essential part of natural selection and therefore of any population of living organisms. In other words individual organisms of the same kind are necessarily more variable than machines or materials and cannot be represented equally well by generic explicit models. We might therefore expect hybrid approaches, which include biological search algorithms in addition to models, to play a larger role in the pharmacological control of living systems.

ACKNOWLEDGMENTS

This work was supported by National Science Foundation grant 0829891.

18. Mayer, R.J., Targeted therapy for advanced colorectal cancer--more is not always better. N Engl J Med, 2009. 360(6): 623-5.
19. Tol, J., M. Koopman, A. Cats, C.J. Rodenburg, G.J. Creemers, J.G. Schrama, F.L. Erdkamp, A.H. Vos, C.J. van Groeningen, H.A. Sinnige, D.J. Richel, E.E. Voest, J.R. Dijkstra, M.E. Vink-Borger, N.F. Antonini, L. Mol, J.H. van Krieken, O. Dalesio, and C.J. Punt, Chemotherapy, bevacizumab, and cetuximab in metastatic colorectal cancer. N Engl J Med, 2009. 360(6): 563-72.
20. Decker, S. and E.A. Sausville, Preclinical modeling of combination treatments: fantasy or requirement? Ann N Y Acad Sci, 2005. 1059: 61-9.
21. DeVita, V.T., S. Hellman, and S.A. Rosenberg, *Cancer, principles & practice of oncology*. 7th ed. 2005, Philadelphia, PA: Lippincott Williams & Wilkins. lxxv, 2898 p.
22. Mendelsohn, J. and G. Powis, From Bench to Bedside with Targeted Therapies. Pp. 521-530. in J. Mendelsohn, et al., ed. The molecular basis of cancer, 2008, Saunders/Elsevier: Philadelphia, PA.
23. Mongan, J.J., T.G. Ferris, and T.H. Lee, Options for slowing the growth of health care costs. N Engl J Med, 2008. 358(14): 1509-14.
24. Wadman, M., The right combination. Nature, 2006. 439: 390-391.
25. Greco, W.R., G. Bravo, and J.C. Parsons, The search for synergy: a critical review from a response surface perspective. Pharmacol Rev, 1995. 47(2): 331-85.
26. Te Dorsthorst, D.T., P.E. Verweij, J. Meletiadis, M. Bergervoet, N.C. Punt, J.F. Meis, and J.W. Mouton, In vitro interaction of flucytosine combined with amphotericin B or fluconazole against thirty-five yeast isolates determined by both the fractional inhibitory concentration index and the response surface approach. Antimicrob Agents Chemother, 2002. 46(9): 2982-9.
27. Borisy, A.A., P.J. Elliott, N.W. Hurst, M.S. Lee, J. Lehar, E.R. Price, G. Serbedzija, G.R. Zimmermann, M.A. Foley, B.R. Stockwell, and C.T. Keith, Systematic discovery of multicomponent therapeutics. Proc Natl Acad Sci U S A, 2003. 100(13): 7977-82.
28. Lamb, J., E.D. Crawford, D. Peck, J.W. Modell, I.C. Blat, M.J. Wrobel, J. Lerner, J.P. Brunet, A. Subramanian, K.N. Ross, M. Reich, H. Hieronymus, G. Wei, S.A. Armstrong, S.J. Haggarty, P.A. Clemons, R. Wei, S.A. Carr, E.S. Lander, and T.R. Golub, The Connectivity Map: using gene-expression signatures to connect small molecules, genes, and disease. Science, 2006. 313(5795): 1929-35.
29. Janes, K.A., J.G. Albeck, S. Gaudet, P.K. Sorger, D.A. Lauffenburger, and M.B. Yaffe, A systems model of signaling identifies a molecular basis set for cytokine-induced apoptosis. Science, 2005. 310(5754): 1646-53.
30. Potti, A., H.K. Dressman, A. Bild, R.F. Riedel, G. Chan, R. Sayer, J. Cragun, H. Cottrill, M.J. Kelley, R. Petersen, D. Harpole, J. Marks, A. Berchuck, G.S. Ginsburg, P. Febbo, J. Lancaster, and J.R. Nevins, Genomic signatures to guide the use of chemotherapeutics. Nat Med, 2006. 12(11): 1294-300.
31. Bechhoefer, J., Feedback for physicists: A tutorial essay on control. Reviews of Modern Physics, 2005. 77: 783-836.
32. Araujo, R.P., E.F. Petricoin, and L.A. Liotta, A mathematical model of combination therapy using the EGFR signaling network. Biosystems, 2005. 80(1): 57-69.